\documentclass[fleqn,10pt]{wlscirep}
\title{Two Types of  Discontinuous Percolation Transitions in Cluster Merging Processes}

\author[1]{Y.S. Cho}
\author[1, *]{B. Kahng}
\affil[1]{Center for Complex Systems Studies and CTP, Department of Physics and Astronomy, Seoul National University, Seoul 151-747, Korea}

\affil[*]{bkahng@snu.ac.kr}

\begin{abstract}
Percolation is a paradigmatic model in disordered systems and has been applied to various natural phenomena. The percolation transition is known as one of the most robust continuous transitions. However, recent extensive studies have revealed that a few models exhibit a discontinuous percolation transition (DPT) in cluster merging processes. Unlike the case of continuous transitions, understanding the nature of discontinuous phase transitions requires a detailed study of the system at hand, which has not been undertaken yet for DPTs. Here we examine the cluster size distribution immediately before an abrupt increase in the order parameter of DPT models and find that DPTs  induced by cluster merging kinetics can be classified into two types. Moreover, the type of DPT can be determined by the key characteristic of whether the cluster kinetic rule is homogeneous with respect to the cluster sizes. We also establish the necessary conditions for each type of DPT, which can be used effectively when the discontinuity of the order parameter is ambiguous, as in the explosive percolation model.  
\end{abstract}
\begin{document}

\flushbottom
\maketitle
%
%
\thispagestyle{empty}


\section*{Introduction}

The percolation transition (PT)~\cite{stauffer}, the emergence of a macroscopic-scale cluster at a finite threshold, has played a central role as a model for metal--insulator and sol--gel \cite{gelation} transitions in physical systems as well as the spread of disease epidemics \cite{epidemic} and opinion formation in complex systems. The ordinary percolation model and many models based on it exhibit continuous transitions as a function of increasing occupation probability. Recently, however, a great interest in discontinuous percolation transitions (DPTs) has been sparked by the explosive percolation model~\cite{explosive} and the cascading failure model in interdependent networks~\cite{havlin, baxter} because of their potential applications to real-world phenomena such as large-scale blackouts in power grid systems and pandemics~\cite{epl}. The explosive percolation model was an attempt to generate a DPT in cluster merging (CM) processes~\cite{explosive, ziff, scalefree, ziff2, friedman,condition,local, multi}, in which clusters are formed as links are added between two unconnected nodes following a given rule. However, recent extensive research~\cite{cont, riordan, lee, nagler} shows that the explosive percolation transition in a random graph is continuous in the thermodynamic limit. This result has reinforced the robustness of continuous PTs in CM processes. 
Along with extensive studies on explosive percolation, a few models exhibiting DPTs in CM processes have been introduced. However, the patterns of DTP that they exhibit are not of the same type, which suggests that further studies are necessary for understanding the mechanism underlying such patterns. In this paper, we classify the patterns into two types and clarify the underlying mechanisms for each type of DPT.

\begin{figure}[t]
\includegraphics[width=0.8\linewidth]{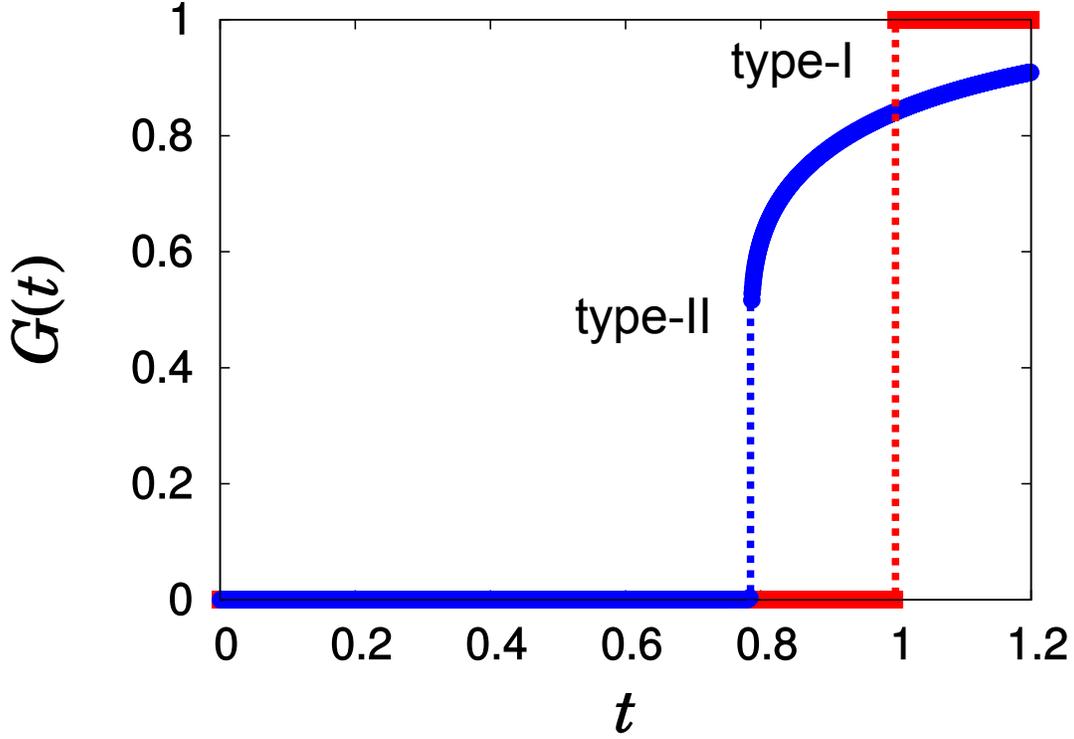}
\centering
\caption{{\bf Schematic diagram of two types of DPTs in CM processes.} $\Delta G=1$ at $t_c = 1$ for type-I, and $\Delta G<1$ at $t_c < 1$ for type-II.
}\label{twotypes_Gt}
\end{figure}

We consider a CM dynamics with $N$ nodes of size one at the beginning. At each time step, an edge is added between two nodes selected according to a given dynamic rule. 
Then, CM kinetics occurs when the two nodes were selected from different clusters. The number of edges added to the system at a certain time step divided by the system size 
$N$ is defined as the time $t$, which serves as a control parameter in PTs. As time passes, the fraction of nodes belonging to the largest cluster in the system, denoted as $G(t)$, 
increases from zero. In the thermodynamic limit $N\to \infty$, $G(t)$, called the order parameter, exhibits a phase transition from zero to $\it O$(1) at a critical point $t_c$. 
Two types of DPTs are possible, as depicted in Fig.~\ref{twotypes_Gt}. For type-I DPTs, the order parameter $G(t)$ increases dramatically with infinite slope all the way to unity at $t_c=1$, whereas 
for type-II DPTs, it also increases similarly but up to a finite value $G(t_c^+) < 1$ at a critical point $t_c < 1$, after which it gradually increases to unity.

The pattern of type-I DPTs in CM processes can be found in various models such as a random aggregation model following the Smoluchowski coagulation equation with reaction kernel $K_{ij}\sim (ij)^{\omega}$ with $0 \le \omega < 0.5$~\cite{cluster}, the Gaussian model~\cite{gaussian}, the avoiding-a-spanning cluster~\cite{science, bridge}, and so on~\cite{makse}. 
The type-II DPT in CM processes can be found in a limited number of mathematical models~\cite{bfw, hierarchical, half}. It would be more interesting to investigate the origin of 
type-II DPTs because this type of DPT can occur in other models, for example, the $k$-core percolation model~\cite{kcore1-1,kcore1-2, kcore2, kcore3}, discontinuous synchronization model~\cite{sync, sync2}, and jamming transition model ~\cite{jamming}.

\section*{Results}


\begin{itemize}
\item Necessary conditions for two types of discontinuous percolation transitions
\end{itemize}

\begin{figure}[t]
\includegraphics[width=1.0\linewidth]{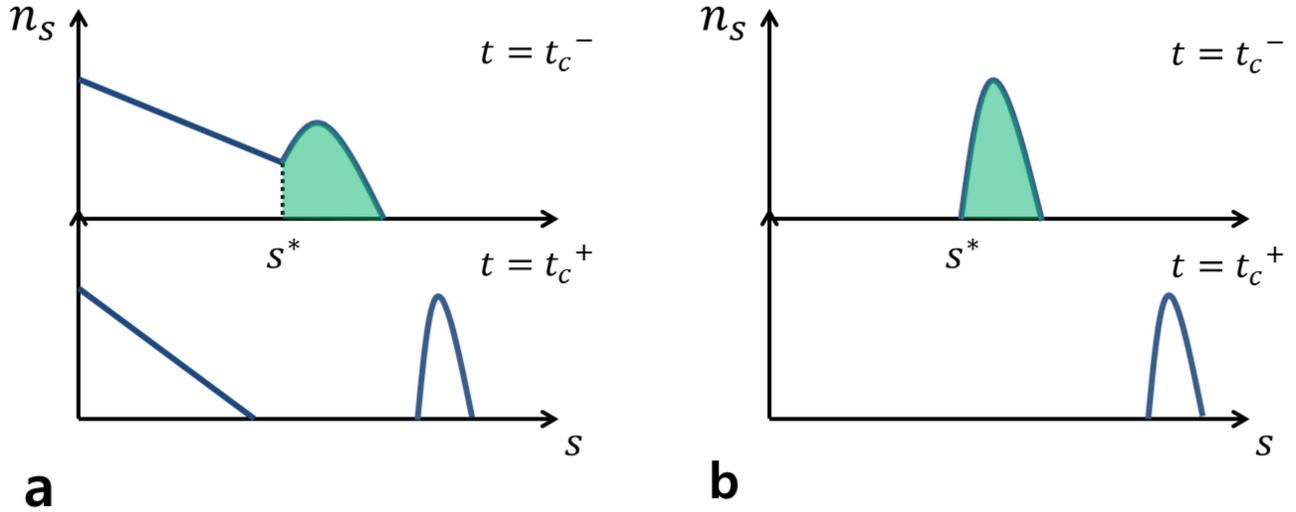}
\centering
\caption{{\bf Schematic illustrations of the cluster size distributions for two types of DPTs.} Schematic illustrations of the cluster size distribution at $t_c^-$ and $t_c^+$ for (a) type-II and (b) type-I are depicted. The number of clusters at $t_c^-$ in the power-law 
region $[1,s^*]$ is ${\it O}(N)$ for (a) and ${\it o}(N)$ for (b).
}\label{twotypescdist}
\end{figure}

Here, we show that the two types of DPTs have different origins. To uncover those origins, we examine the cluster size distributions immediately before and after the percolation threshold, denoted as $t_c^-$ and $t_c^+$, respectively, and defined later in the Methods. To induce a type-II DPT, it is necessary that the clusters at $t_c^-$ are heterogeneous in size, ranging from small cluster sizes to large ones.  Among those clusters, primarily large clusters merge to create a macroscopic-scale giant cluster during a short time interval $[t_c^{-},t_c^{+}]$. Beyond $t_c^{+}$, most of the merging is caused by remnant small clusters, the number of which is still $\it{O}(N)$, which mainly join the giant cluster. On the other hand, for a type-I DPT, at $t_c^{-}$, the remaining clusters are mainly homogeneous with mesoscopic-scale size, and they merge during the interval $[t_c^{-},t_c^{+}]$ to create a macroscopic-scale giant cluster. A schematic comparison of these kinetics between DPTs of types II and I is shown in Fig.~\ref{twotypescdist}. 

\begin{figure}[t]
\includegraphics[width=1.0\linewidth]{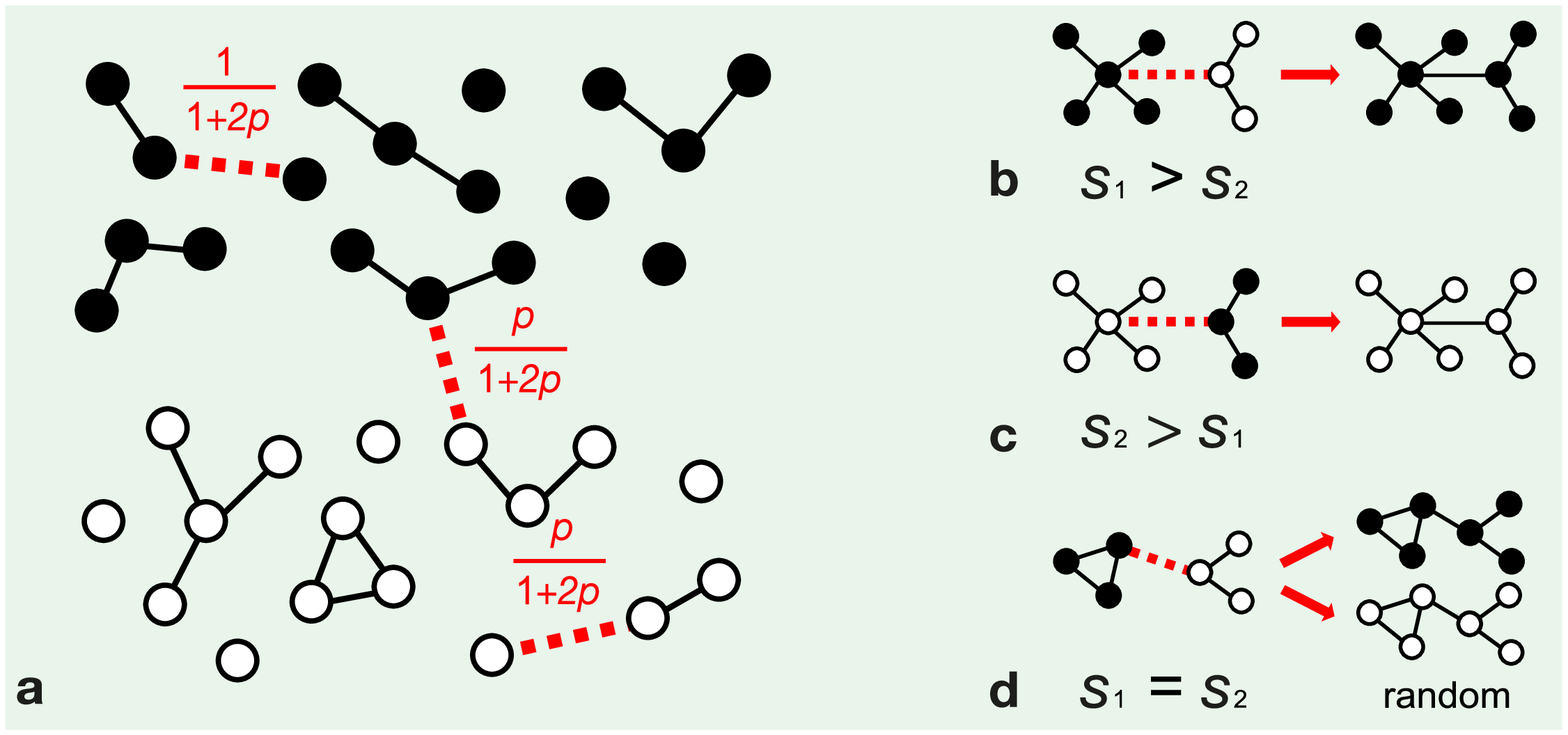}
\centering
\caption{{\bf Schematic illustrations of the dynamic rule of the TCA model.} (a) There are three types of CM processes, each of which depends on the species of merging clusters. The probabilities for each case are given in the figure. 
}\label{model_schematic}
\end{figure}

To quantify the origin, we propose the necessary conditions for each type of DPT as follows. Here $n_s(t)$ denotes the  number of $s$-size clusters divided by $N$, which changes with time. 

\begin{itemize}
\item[I)]Necessary condition for type-II DPT: At $t_c^{-}$, at least one characteristic cluster size $s^{*}>1$ has to exist, which fulfills the following conditions in the thermodynamic limit:
I-i) $\sum_{s=s^{*}}^{\infty}n_s(t_c^{-}) \rightarrow 0$,  
I-ii) $\sum_{s=1}^{\infty}n_s(t_c^{-}) \sim {\it O}(1)$, and 
I-iii) $\sum_{s=s^{*}}^{\infty}sn_s(t_c^{-}) \rightarrow r$ $(0<r<1)$.

\item[II)]Necessary condition for type-I DPT: At $t_c^-$, at least one characteristic cluster size $s^*>0$ has to exist, which fulfills the following conditions in the thermodynamic limit: I-i) $\sum_{s=s^*}^{\infty}n_s(t_c^-)\rightarrow 0$, and II-ii) $\sum_{s=s^*}^{\infty}sn_s(t_c^-) \rightarrow 1$.
\end{itemize}
The derivations of the two necessary conditions are presented in the Methods.



\begin{itemize}
\item Two-species cluster aggregation model
\end{itemize}
We introduce a cluster aggregation model that exhibits both type-I and -II DPTs as the model parameter changes. The dynamic rule is depicted schematically in Fig.~\ref{model_schematic}. For this model, which is referred to as the two-species cluster aggregation (TCA) model, we start with $N$ isolated nodes, half of which are colored black and the other half of which are white. The color may represent opinion, for example, the left- and right-wing positions on a political issue. According to the dynamic rule below, all nodes in the same cluster have the same color: either black or white. At each time step, we first select one case among the three possible combinations, (black, black), (black, white), or (white, white), with probabilities $1/(1+2p)$, $p/(1+2p)$, and $p/(1+2p)$, respectively, where $p$ is a model parameter in the range $0< p \le 1$. Second, two clusters 
are selected following the colors selected but independently of the cluster sizes. 
Finally, two nodes---one from each cluster---are selected randomly and connected, which causes the two clusters to merge. If the two selected clusters are the same, then two distinct nodes from that cluster are connected.

\begin{figure*}[t!]
\includegraphics[width=1.0\linewidth]{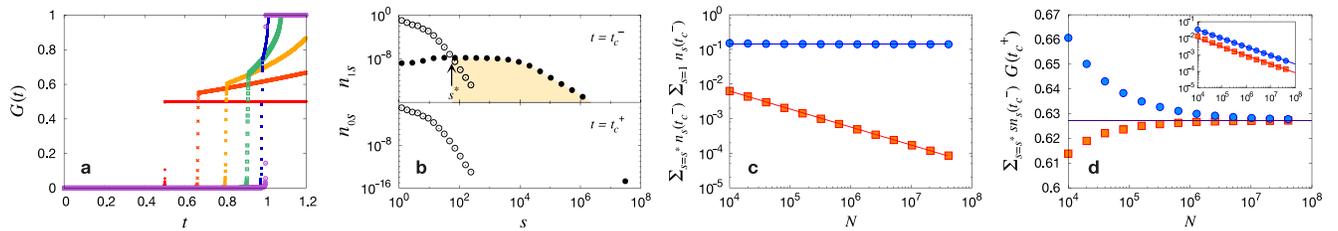}
\centering
\caption{{\bf Numerical tests of necessary conditions for type-II DPT in TCA model.} (a) $G(t)$ vs. $t$ in the TCA model for various values of $p$ for a system size of $N=10^{5}$. From left to right, $p=0, 0.2, 0.4, 0.6, 0.8$, and $1.0$. (b) The cluster size distributions of black clusters $n_{0s}(\bullet)$ and white clusters $n_{1s}(\circ)$ at $t_c^{-}$ (upper panel) and $t_c^{+}$ (lower panel) for $N=2^{12}\times 10^{4}$.  (c) $\sum_{s=s^{*}}^{\infty}n_s(t_c^{-})(\square)$ and $\sum_{s=1}^{\infty}n_s(t_c^{-})(\circ)$ vs. $N$. (d) $\sum_{s=s^{*}}^{\infty}sn_s(t_c^{-})(\square)$ and $G(t_c^{+})(\circ)$ vs. $N$. The two data sets converge to the value $y_0 \approx 0.63$. Inset: $y_0 - \sum_{s=s^{*}}^{\infty}sn_s(t_{c}^-)(\square)$ and $G(t_c^+)-y_0(\circ)$ vs. $N$, respectively. The slopes of the guidelines for $(\square)$ in (c) and the inset of (d) are equal to $-0.52$. The data sets for (b), (c), and (d) are obtained for $p=0.5$.}\label{TCA_necessary}
 \end{figure*}

The colors of all the nodes in the resulting merged cluster are updated according to the following rule: if the colors of the two clusters are the same, there is no change. However, if the colors are different, then the colors of all the nodes in the smaller cluster are changed to that of the larger cluster. This change may represent opinion formation following the so-called majority rule. If the clusters have the same size but different colors, then either color is picked with equal probability. We numerically show that if $0<p<1$, the PT is discontinuous and occurs at a finite threshold, $t_c < 1$, and if $p=1$,  $t_c=1$ [Fig.~\ref{TCA_necessary}(a)]. We note that if $0<p<1$, the symmetry between different species in the dynamic rule is broken; if $p=1$, the symmetry is preserved. 

The dynamic rule, particularly in the process of updating the color of nodes, can be modified in several ways.  Nevertheless, the overall behavior of the DPT does not change significantly. To facilitate an analytic solution, we modify the dynamic rule as follows: When the colors of the two selected clusters are different, we take black regardless of the cluster size, i.e., without following the majority rule. This modification enables us to set up a coupled Smoluchowski coagulation equation and, consequently, to analytically understand the evolution of a large cluster. When $0<p<0.5$, the PT is discontinuous at a 
finite threshold $t_c<1$ and $\Delta G<1$ (type-II DPT). When $p \geq 0.5$, the PT is also discontinuous, but at
$t_c=1$ and $\Delta G=1$ (type-I DPT). A detailed derivation is presented in the Methods.

\begin{figure*}
\includegraphics[width=0.9\linewidth]{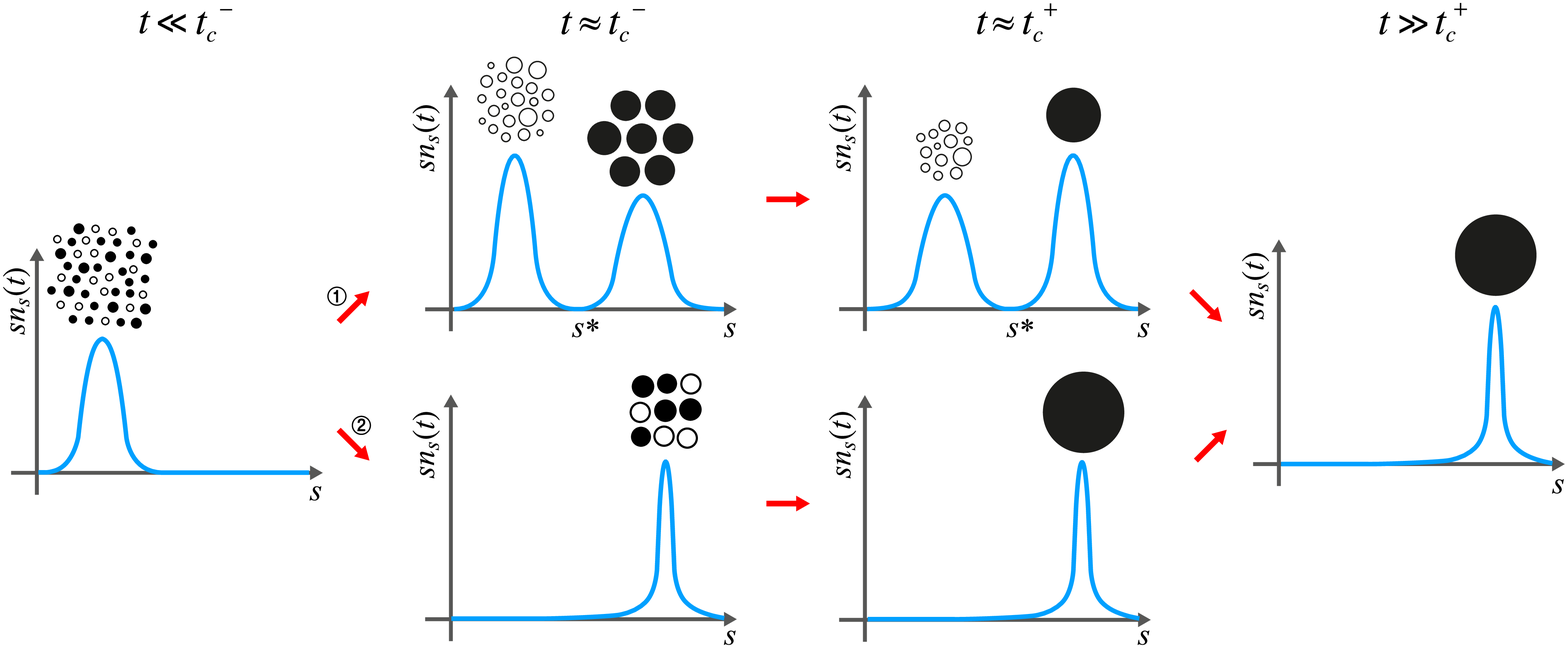}
\centering
\caption{{\bf Schematic illustration of symmetry-preserving (-breaking) dynamics.} Schematic illustration of the formation of type-I and -II DPTs  in CM processes  through \textcircled{\footnotesize 1} upper pathway and \textcircled{\footnotesize 2} lower pathway, 
respectively.}\label{symmetry}
\end{figure*}

\begin{figure}
\includegraphics[width=0.6\linewidth]{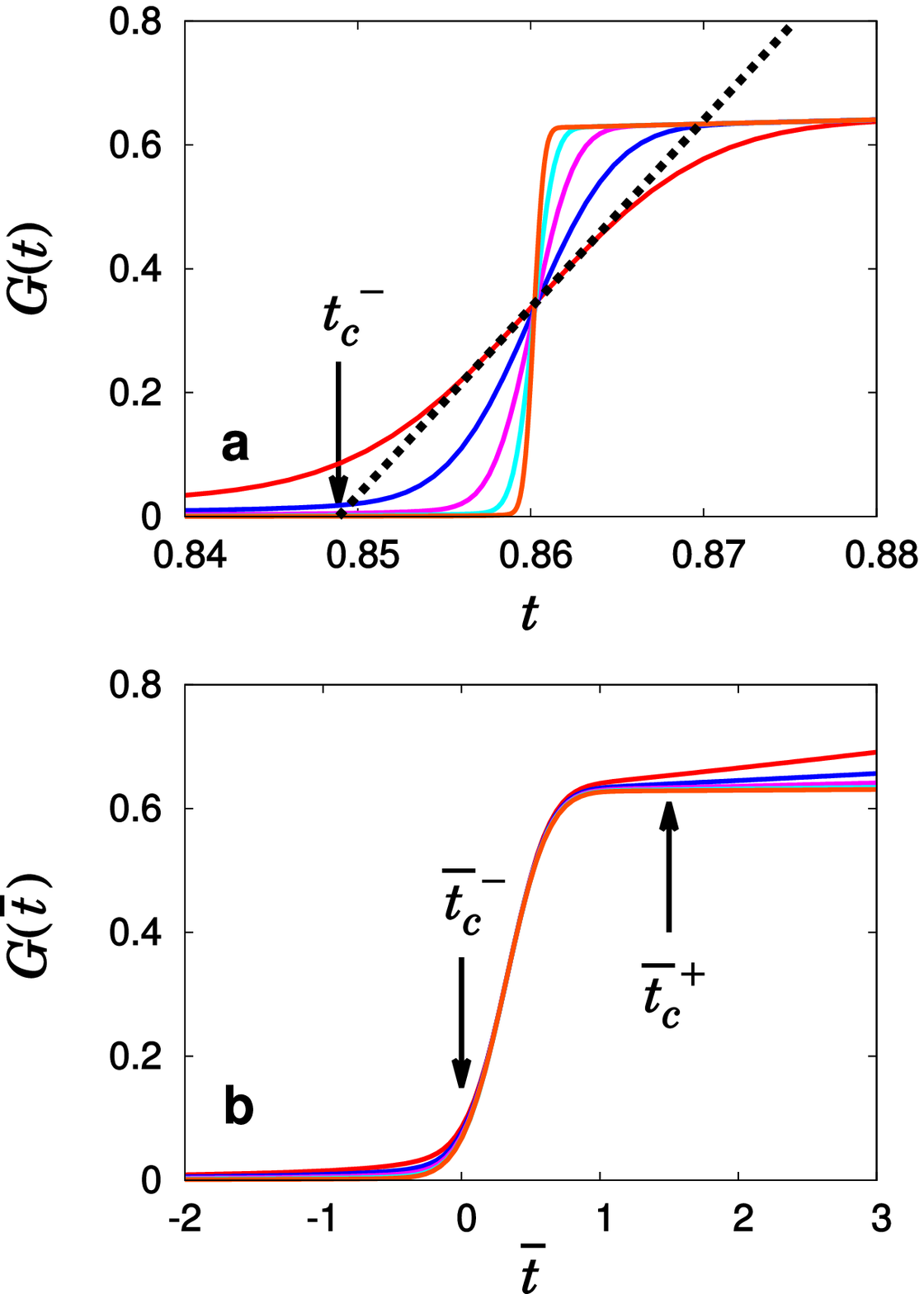}
\centering
\caption{{\bf $t_c^-$ and $t_c^+$ used for numerical tests.} $G_N(t)$ vs. $t$ for the TCA model with $p=0.5$ for different system sizes, $N/10^4 = 1, 4, 16, 64$, and $256$. 
(a) We draw a tangent at the time at which the slope $dG_N(t)/dt|_{\textrm{max}}$ becomes 
maximum, which is almost independent of $N$ and denoted as $t_c$. The $t$ intercept of the tangent 
of the curve $G_N(t)$ is denoted as $t_c^-(N)$. As the system size $N$ is increased, the slope $dG_N(t)/dt|_{\textrm{max}}$ increases. (b) We plot $G_N({\bar t})$ of different system sizes vs. a rescaled time 
as $\bar{t} \equiv (t-t_c^-(N))dG_N(t)/dt|_{\textrm{max}}$. 
Then, the $\bar t$ intercept of the tangent of the curve $G_N({\bar t})$ is denoted as ${\bar t}_c^-=0$, which is independent of $N$. Next, we take $\bar{t_c}^+$ as a crossover point from which $G({\bar t})$ 
begins to grow gradually. Then, $t_c^+(N) \equiv t_c^-(N) + {\bar t}_c^+(dG_N(t)/dt|_{\textrm max})^{-1}$.
}\label{tcminus_tcplus}
\end{figure}

\begin{itemize}
\item Numerical tests and symmetry-preserving (-breaking) dynamics
\end{itemize}

 Here we test the necessary conditions for the TCA model and clarify the origin of the type-II DPT.  For this purpose, we plot the cluster size distribution for the TCA model at $t_c^-$ and $t_c^+$ in Fig.~\ref{TCA_necessary}(b).
At $t_c^{-}$, the size distributions of the white and black clusters decay exponentially in the asymptotic region. However, the size distribution of the black clusters is extended to a larger region owing to the symmetry-breaking properties of the dynamic rule. The nodes belonging to the extended (shaded) region correspond to the powder keg referred to in previous studies ~\cite{friedman, condition, riordan}. The combined cluster size distribution exhibits crossover behavior from the region primarily composed of white clusters to that primarily composed of black clusters across a characteristic size, which we denote as $s^{*}$. This segregation is induced by the symmetry-breaking dynamic rule: Merging occurs with a higher probability between black clusters than between other types of clusters. Thus, black clusters grow more rapidly and belong to the region $s> s^*$. The cluster size distribution at $t_c^{+}$, when the dramatically increasing order parameter $G(t)$ changes to a gradually increasing $G(t)$, is shown in the lower panel of Fig.~\ref{TCA_necessary}(b). The difference between the two figures shows that during the interval $t_c^{+}-t_c^{-}$, almost all the black clusters aggregate to form a large cluster, and a small number of white clusters merge with large black clusters as black clusters. This microscopic understanding of the mechanism of a type-II DPT is schematically illustrated in Fig.~\ref{symmetry}. This origin can also be observed in other models such as the so-called Bohman--Frieze--Wormald (BFW) model \cite{bfw} and a half-restricted process model \cite{half}, which are shown in the supplementary information. On the basis of these numerical results for the merging processes, we made the assumption stated previously when the necessary conditions were set up.  

We numerically confirm the necessary conditions that the number of clusters of size $s>s^{*}$ at $t_c^{-}$ is sub-extensive [condition I-i)]. The total number of clusters over the entire range of $s$ is, however, extensive to $N$ [condition I-ii)] [Fig.~\ref{TCA_necessary}(c)], which is needed for a gradual increase of the order parameter beyond $t_c^{+}$. Next, we measure the number of nodes belonging to clusters of sizes $s>s^{*}$, finding that the order parameter converges to a finite value $r\approx 0.63 < 1$ as the system size is increased. The nodes belonging to this region become the elements of a macroscopic-scale giant cluster, as can be seen for large $N$ cases [Fig.~\ref{TCA_necessary}(d)]. Numerical testing of the necessary conditions is performed for other models such as 
the BFW model~\cite{bfw}, the half-restricted process model~\cite{half}, and the ordinary percolation model in a hierarchical network with long-range connection~\cite{hierarchical}.
The details are presented in the supplementary information.

%
%
%
%

\section*{Discussion}


We investigated the origins of the two types of DPTs in CM processes and derived the necessary conditions for them. 
Our derivation is similar to the picture proposed by Friedman and Landsberg~\cite{friedman}, in which the occurrence of an abrupt PT is determined by the number of the clusters in the powder keg region with $s> s^*$. They set the characteristic size as $s^*\sim N^{1-\beta}$ with $\beta < 1$.  Then, $\Delta t < N^{\beta-1}$, which is reduced to zero in the limit $N\to \infty$. This criterion is the same as condition I-i) we obtained here. On the other hand, the authors of \cite{friedman} did not classify the necessary conditions for a type-I or -II DPT separately.  In a similar way, Hooyberghs and Schaeybroeck~\cite{condition} proposed another criterion for a DPT, which is again limited to our necessary condition for a type-I DPT. 

We have also introduced an analytically solvable model in which two species of clusters evolve through CM processes under the symmetry-breaking rule and showed that this symmetry breaking dynamics generates a type-II DPT.  This phenomena can also be found in a model for synchronization transition \cite{sync,sync2}. The detail is presented in SI.   

We remark that the origin of the type-II DPT in CM processes differs from that of DPTs driven by the cascading failure dynamics in interdependent networks~\cite{havlin} or in the $k$-core percolation model~\cite{kcore2}. The cluster size distribution at $t_c^-$ for the latter case does not resemble that in the former case. Thus, the necessary conditions we studied cannot be applied to the latter case. In addition, when a type-II DPT is induced by the hierarchical structure as in \cite{hierarchical}, even though our necessary conditions were found to be valid, it is not clear yet whether the DPT originates from the symmetry-breaking kinetics.

\section*{Methods}


\begin{itemize}
\item Numerical testing
\end{itemize}
It is necessary to use the appropriate times $t_c^-$ and $t_c^+$. In Fig.~\ref{tcminus_tcplus}, we illustrate how to take $t_c^-(N)$ and $t_c^+(N)$ in numerical tests of the necessary conditions. We used more than $O(10^{11}/N)$ configurations for all numerical analyses. \\

\begin{itemize}
\item Derivation of the necessary conditions
\end{itemize}
 To derive the necessary conditions, we suppose the extreme case, in which CM dynamics occurs only between clusters of size $s>s^*$ during a short time interval within $[t_c^{-},t_c^{+}]$. In this case, when intercluster links are added, the order parameter can increase the most rapidly. The number of links to connect all those clusters divided by $N$ is $\sum_{s=s^*}^{\infty}n_s(t_c^{-})$, which is equivalent to $\Delta t \equiv t_c^{+}-t_c^{-}$. 

First, we consider a type-II DPT. To verify condition I-i), we use the fact that if $\sum_{s=s^*}^{\infty}n_s(t_c^{-}) \rightarrow 0$ in the limit $N \rightarrow \infty$, then $\Delta t \rightarrow 0$. During this interval, because the order parameter increases as much as ${\it O}(1)$, the PT is discontinuous. Thus, condition I-i) provides a necessary condition for a discontinuous PT. To verify condition I-ii), we consider the inequality $1-\sum_{s=1}^{\infty}n_s(t_c^{-}) \leq t_c^{-}$, which comes from the fact that the number of links added up to $t_c^-$ is larger than (or equal to) the number of CM events. The equality holds when the model disallows the attachment of intracluster links. In general, when $\sum_{s=1}^{\infty}n_s(t_c^-)$ goes to zero, $t_c^{-} \geq1$ in the thermodynamic limit. Condition I-ii), $\lim_{N \rightarrow \infty}\sum_{s=1}^{\infty}n_s(t_c^-) \sim {\it O}(1)$, provides a necessary condition for the transition point to be $t_c^{-}<1$. 
Next, let us define $r=\sum_{s=s^*}^{\infty}sn_s(t_c^-)$, which corresponds to the size of the powder keg in \cite{friedman}. 
Then, $r=1-\sum_{s=1}^{s^*-1}sn_s(t_c^-)$. This quantity satisfies the following inequality:
$r\leq1-\sum_{s=1}^{s^*-1}n_s(t_c^-)=1-\sum_{s=1}^{\infty}n_s(t_c^-)+\sum_{s=s^*}^{\infty}n_s(t_c^-)$. When conditions I-i) and I-ii) hold, 
$r\leq1-{\it O}(1)$. Thus, $r<1$. Condition I-iii) is needed to exclude the case $r=0$ for a continuous 
transition. Thus, conditions I-i), I-ii), and I-iii) are all needed for a type-II DPT.  

We now consider a type-I DPT. Condition I-i) suggests that $\Delta t \rightarrow 0$ in the thermodynamic limit.
Condition II-ii) suggests that $G(t_c^+) \rightarrow 1$. Then, using the inequality
$1-\sum_{s=1}^{\infty}n_s(t_c^-) \leq t_c^{-}$, one can obtain $t_c^{-} \geq 1$ in the thermodynamic limit. Thus, the percolation threshold is positioned at $t_c \geq 1$. We remark that this necessary condition 
for a type-I DPT in CM processes is equivalent to $\lim_{N \rightarrow \infty}\sum_{s=1}^{\infty}n_s(t_c^-)=0$.\\

\begin{itemize}
\item Analytic calculation of the solvable two-species cluster aggregation model
\end{itemize}
 Let $n_{0s}(t)$ and $n_{1s}(t)$ be the numbers of $s$-size black and white clusters per node, respectively, at time step $t$. The rate equations of the two quantities are written as
\begin{equation}
\\\\\ \frac{dn_{0s}}{dt}=\frac{1}{1+2p}\left(\sum_{i=1,j=1}^{\infty}\frac{n_{0i}}{c_0}\frac{n_{0j}}{c_0}\delta_{i+j,s}-2\frac{n_{0s}}{c_0}\right)
+\frac{p}{1+2p}\left(\sum_{i=1,j=1}^{\infty}\frac{n_{0i}}{c_0}\frac{n_{1j}}{c_1}\delta_{i+j,s}-\frac{n_{0s}}{c_0}\right),\label{rate1}
\end{equation}
\begin{equation}
\\\\\ \frac{dn_{1s}}{dt}=\frac{p}{1+2p}\left(\sum_{i=1,j=1}^{\infty}\frac{n_{1i}}{c_1}\frac{n_{1j}}{c_1}\delta_{i+j,s}-2\frac{n_{1s}}{c_1}\right)
-\frac{p}{1+2p}\frac{n_{1s}}{c_1},\label{rate2}
\end{equation}
where $c_0 = \sum_{s=1}^{\infty}n_{0s}(t)$ and $c_1=\sum_{s=1}^{\infty}n_{1s}(t)$ are the number of finite black and white clusters per node at time $t$ in the system, respectively. Next, we define the generating functions $f(z,t)=\sum_{s=1}^{\infty}n_{0s}(t)z^s$ and $g(z,t)=\sum_{s=1}^{\infty}n_{1s}(t)z^s$, where the summation runs over finite clusters. As a result, the rate equations (\ref{rate1}) and (\ref{rate2}) are changed to
\begin{equation}
\\\\\ \frac{df(z,t)}{dt}=\frac{1}{1+2p}\left(\frac{f^2}{c_0^2}-\frac{2f}{c_0}\right)+\frac{p}{1+2p}
\left(\frac{fg}{c_0c_1}-\frac{f}{c_0}\right),\label{rate3}
\end{equation}
\begin{equation}
\\\\\ \frac{dg(z,t)}{dt}=\frac{p}{1+2p}\left(\frac{g^2}{c_1^2}-\frac{2g}{c_1}\right)-\frac{p}{1+2p}
\frac{g}{c_1}. \label{rate4}
\end{equation}

Using $c_0(t)=f(1,t)$ and $c_1(t)=g(1,t)$, we obtain
$c_0(t)=1/2-t/(1+2p)$ and $c_1(t)=1/2-2pt/(1+2p)$. 
When $0<p<0.5$, the percolation threshold can be obtained by setting $c_0(t_c)=0$ but $c_1(t_c)>0$, 
because $c_0(t)$ decreases more rapidly than $c_1(t)$. Thus, $t_c=1/2+p$, and a large black cluster 
emerges at $t_c$. The size of the jump in the order parameter at $t_c$ can be obtained using the 
formula $\Delta G=1-f'(1,t_c)-g'(1,t_c)$, which reduces to $\Delta G=1-g'(1,t_c)$, 
because $f'(1,t_c)=0$. Thus, the jump in the order parameter is determined to be $\Delta G=1-\sqrt{1-2p}/2$.
The PT is discontinuous at a finite threshold $t_c<1$ and $\Delta G<1$ (type-II DPT).

When $p \geq 0.5$, because $c_1(t)$ decreases more rapidly than $c_0(t)$, the percolation threshold can be obtained
using $c_1(t_c)=0$ and $c_0(t_c)>0$. Thus, $t_c=\frac{1}{2}+\frac{1}{4p}$. The size of the jump in the order parameter can be 
obtained using the formula $\Delta G=1-f'(1,t_c)$. However, $f'(1,t_c)=1$ and $f'(1,t)=1$ even for $t<1$. Thus, $\Delta G = 0$ for $t<1$. When $t>1$, $f'(1,t)=0$. 
Thus, the order parameter behaves as $\Delta G=1$ for $t>1$, and the threshold $t_c=1$ (type-I DPT).
These analytic results are checked numerically in the supplementary information.



\section*{Acknowledgements (not compulsory)}

This work was supported by NRF grants (Grant Nos. 2010-0015066 \& 2014R1A3A2069005) (BK) and the Global Frontier Program (YSC).


\section*{Author contributions statement}

B.K. wrote the main manuscript text, and Y.S.C. performed all the simulations. Both authors analyzed all the data. 


\section*{Additional information}

Competing financial interests: The authors declare no competing financial interests.

%

%
%

\end{document}